\documentclass[preprint,showpacs,preprintnumbers,amsmath,amssymb,nofootinbib]{revtex4}

\usepackage{epsfig}
\usepackage{graphicx}
\usepackage{dcolumn}
\usepackage{bm}

\def\DESepsf(#1 width #2){\epsfxsize=#2 \epsfbox{#1}}

\begin{document}

\preprint{November, 2005}

\title{Extracting the unitarity angle $\gamma$ in $B_s\to D^0 h^0,\,\overline D {}^0 h^0$ Decays}
%

\author{Chun-Khiang Chua}
\affiliation{Institute of Physics, Academia Sinica, Taipei, Taiwan
115, Republic of China}


\begin{abstract}
The recently observed color-suppressed $\overline B{}^0 \to
D^0\pi^0$, $D^0\eta^{(\prime)}$, $D_s^+ K^-$, $D^0\overline K
{}^0$, $D^0 \rho^0$ and $D^0 \omega$ decay modes all have rates
larger than expected. The color-suppressed $B_s\to
D^0\phi,\,\overline D {}^0\phi$ modes, which were suggested for
the extraction of the unitarity angle $\gamma$ in the
Gronau-London method, could be larger than the previous estimation
by one order of magnitude. Several new theoretical clean modes in
$B_s$ decays are suggested for the extraction of $\gamma$. The
proposed $B_s\to D^0 h^0,\,\overline D {}^0h^0$ decay modes with
$h^0=\pi^0,\,\eta,\,\eta',\,\rho^0,\,\omega$ in addition to
$h^0=\phi$ are free from penguin contributions. Their decay rates
can be estimated from the observed color-suppressed $\overline
B^0\to D^0 h^0$ rates through SU(3) symmetry. A combined study of
these $D^0 h^0,\,\overline D {}^0 h^0$ modes in addition to the
$D^0\phi,\,\overline D {}^0\phi$ modes is useful in the extraction
of $\gamma$ in the $B_s$ system without involving
$B_s$--$\overline B_s$ mixing. Since the $b\to u$ and $b\to c$
transitions belong to the same topological diagram, the relative
strong phase is likely to be small. In this case, the $CP$
asymmetries are suppressed and the untagged rates are very useful
in the $\gamma$ extraction.
\end{abstract}

\maketitle


The extraction of the unitarity angle $\gamma\equiv\arg V_{ub}^*$,
where $V$ is the Cabibbo-Kobayashi-Maskawa (CKM) quark mixing
matrix, is important in completing or testing the Standard Model
(SM). Several theoretical clean ways of the weak phase extraction
were proposed using interference effects. At $B$ factories, the
extraction is performed in the $DK$ system, using the interference
effect of $B\to D^0K$ and $\overline D {}^0 K$ decays in $D_{CP}K$
final states, where $D_{CP}$ are the CP eigenstates of $D^0$ and
$\overline D {}^0$ mesons, or to some common $f_{CP} K$, $f_{CP}$
states~\cite{GL,GW,ADS,Dalitz,DalitzBelle0304,Gronau:2004gt}.
Similarly, the color-suppressed $D_{CP}\phi$ mode was also
proposed in the extraction of $\gamma$ in the $B_s$
system~\cite{GL}. An alternative method made use of the
$B_s$--$\overline B_s$ mixing was proposed using color-allowed
$B_s\to D^\pm_s K^\mp$ decays with time-dependent
tagging~\cite{ADK}. Due to the large rate ($10^{-4}$) in the
color-allowed decays, this scenario has been seriously considered
at LHCb~\cite{LHCb}.

It is well known that in the SM, the $\Delta m_{B_s}$ in the $B_s$
system is much larger than the one in the $B_d$ system.
Experimental searches give $\Delta
m_{B_s}>14.5$~ps$^{-1}$~\cite{PDG}. The measurement of the
time-dependent asymmetry in the $B_s$ system is challenging.
Furthermore, the deviation of the recently measured
$\sin2\beta_{\rm eff}$ in penguin-dominated modes from the
$\sin2\beta$ ($\beta\equiv\arg V_{td}^*$) extracted from
charmonium modes may hint at New Physics contributions in the $b
\to s$ transitions~\cite{HFAG,sin2beta}. In this case, the $\Delta
m_{B_s}$ can easily be much larger than the SM expectation (see,
for example~\cite{b2sNP}). Therefore, an extraction of $\gamma$
without relaying on the $B_s$--$\overline B_s$ mixing is
complementary to the $D^\pm_s K^\mp$ program and is indispensable
to the $\gamma$ program in the $B_s$ system.

Although the Gronau-London $D_{CP}\phi$ method~\cite{GL} does not
need time-dependent tagging, its usefulness is questioned by the
smallness of the color-suppressed decay rate, which is estimated
to be as small as $10^{-6}$~\cite{ADK}.
However, color-suppressed $\overline B {}^0\to
D^{(*)0}\pi^0,\,D^0\eta^{(\prime)},\,D^0\omega,\,D^0\rho^0,\,D_s^+
K^-,\,D^0 \overline K{}^0$ decay modes were observed with
branching ratios significantly larger than earlier theoretical
expectations based on naive factorization~\cite{D0h0}.
The large color-suppressed decay rates have attracted much
attention~\cite{D0h0theory1,D0h0theory2,D0h0theory3,D0h0theory4,CH05}.
Similar enhancement in the color-suppressed decay rates in the
$B_s$ system is expected. In particular, the $D^0\phi$ rate is
expected to be larger than the previous estimation.
In addition to the $D\phi$ mode, several other theoretical clean
modes are suggested in this work. The proposed tree $D^0
h^0,\,\overline D {}^0h^0$ decay modes, where
$h^0=\pi^0,\,\eta,\,\eta',\,\rho^0,\,\omega$, in additional to the
$D^0\phi,\,\overline D {}^0\phi$ modes are useful to extract
$\gamma$ without time-dependent tagging. As we shall see later,
the extraction done only with untagged rates can also be useful.

\begin{figure}[t!]
\centerline{\hskip-1.1cm \DESepsf(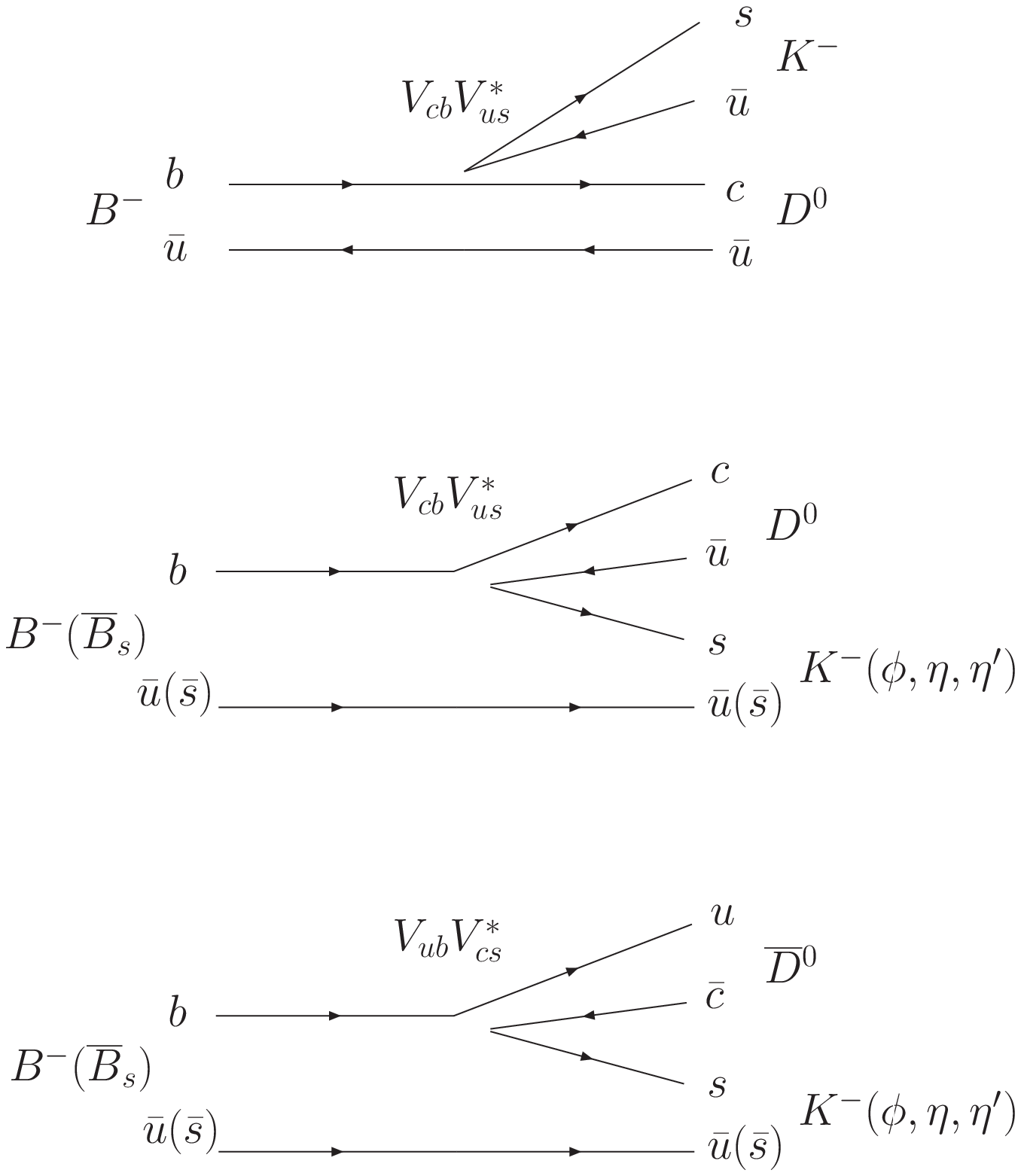 width 9cm)}
\caption{Color-allowed and color-suppressed amplitudes for $B^-\to
D^0 K^-$ decay, and color-suppressed amplitude for the
$B^-(\overline B_s)\to\overline D {}^0 K^-(\phi,\eta,\eta')$
decay.
 }
 \label{fig:DK}
\end{figure}

In this study, the $\gamma$ extraction method is similar to the
$B_d\to D K$ and $B_s\to D\phi$ method. It will be useful to
briefly review the $DK$ method and the present experimental status
at $B$ factories. To be specific, the amplitude ratio $r_B$ and
the strong phase difference $\delta_B$ for the color-allowed
$\overline B^0\to \overline D {}^0 K^-$ and color-suppressed $D^0
K^-$ decays, which are governed by different CKM matrices as
depicted in Fig.~\ref{fig:DK}, are defined as
 \begin{equation}
 r_B=\left|\frac{A(B^-\to \overline D {}^0 K^-)}{A(B^-\to D {}^0
 K^-)}\right|,
 \qquad
 \delta_B=\arg\left[\frac{e^{i\gamma}
  A(B^-\to \overline D {}^0 K^-)}{A(B^-\to D {}^0 K^-)}\right].
 \label{eq:rB}
 \end{equation}
The weak phase $\gamma$ is removed from $A(B^-\to \overline D {}^0
K^-)$ in the $\delta_B$ definition. Since the strong phase
difference arises from that in the color-suppressed and
color-allowed amplitudes, it is expected to be non-vanishing. The
$r_B$ and $\delta_B$ parameters are common to the $\gamma$
determination methods of Gronau-London-Wyler (GLW)~\cite{GL,GW},
Atwood-Dunietz-Soni (ADS)~\cite{ADS} and ``$DK$ Dalitz
plot"~\cite{Dalitz,DalitzBelle0304}, where one exploits the
interference effects of $B^-\to D^0 K^-\to f_{CP} K^-$ and $B^-\to
\overline D {}^0 K^-\to f_{CP}K^-$ amplitudes. Note that the $r_B$
parameter, which governs the strength of interference, is both
color and CKM suppressed, hence hard to measure directly.

Through the $DK$ Dalitz plot method, BaBar and Belle experiments
already find $\gamma=67^\circ\pm 28^\circ\pm13^\circ\pm11^\circ$
and $64^\circ\pm 19^\circ\pm13^\circ\pm11^\circ$,
respectively~\cite{HFAG,gammaDK}, where the last error comes from
modelling of $D$ decay resonances across the Dalitz plot for, e.g.
$D^0 \to K_S\pi^+\pi^-$, and the BaBar measurement includes the
$DK^*$ analysis. Although similar results on $\gamma$ are
obtained, the corresponding $r_B$ values are quite different for
BaBar and Belle. Belle reports $r_B=0.21\pm0.08\pm0.03\pm0.04$ and
$\delta_B=(157\pm 19\pm 11\pm 21)^\circ$, while BaBar gives
$r_B=0.118\pm 0.079\pm0.034^{+0.036}_{-0.034}$ and
$\delta_B=(104\pm 45^{+17 +16}_{-21-24})^\circ$. Note that an
average of $r_B=0.10\pm 0.04$ is found by the UT$_{fit}$ group, by
combining analyses using all three methods~\cite{UTfit}. As the
strength of interference is governed by the size of $r_B$, the
larger error in the $\gamma$ value of BaBar reflects the smallness
of their $r_B$. Given the present experimental situation that
Belle and BaBar have quite different $r_B$ values and that the
critical role it plays in the $\gamma$ extraction, it is important
to compare with a theoretical or phenomenological prediction of
$r_B$. In a recent work, we obtained
$r_B=0.09\pm0.02$~\cite{CH05}. The predicted $r_B$ agrees with the
UT$_{fit}$ extraction~\cite{UTfit} and does not differ much from
the naive factorization expectation. Furthermore, the $r_B$ value
prefers the lower value of the BaBar experiment and disfavors the
Belle result. A similar $r_B$ was found experimentally in the
$DK^*$ analysis~\cite{HFAG,gammaDK}. The smallness of the ratio
$r_B$ would demand larger statistics of data for the $\gamma$
program in the $DK^{(*)}$ system. In fact, the smallness of $r_B$
is precisely the reason that ADS and $DK$ Dalitz methods are
needed in additional to the original GLW method. However, these
methods usually bring in additional uncertainties, such as the
fourth uncertainties in the extracted $\gamma$ value quoted above.

\begin{figure}[t!]
\centerline{\hskip-1.1cm \DESepsf(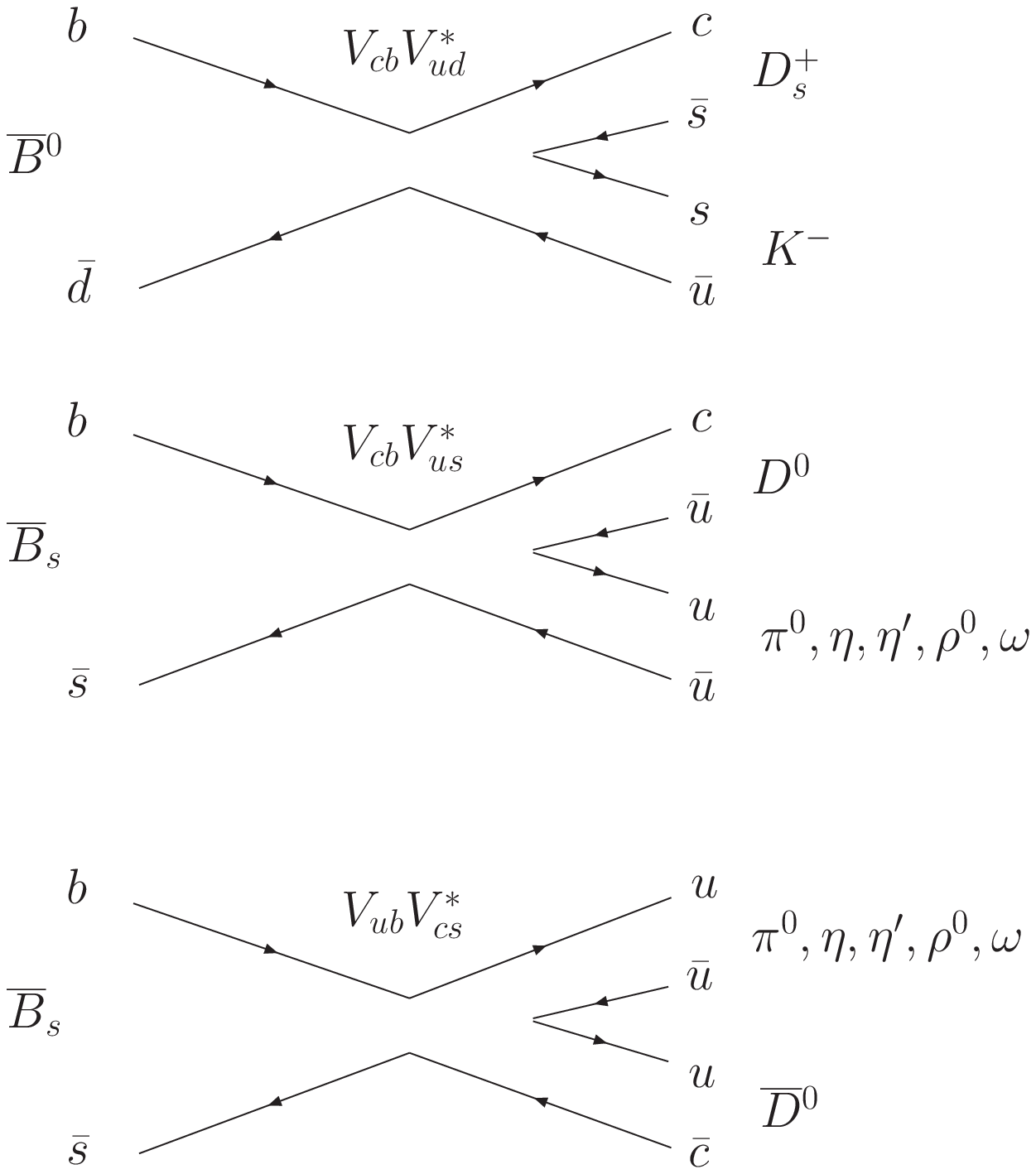 width 9cm)}
\caption{$W$-exchanged amplitudes for $\overline B {}^0\to D_s^+
K^-$ and $\overline B_s\to D^0 h^0$ decays.
 }
 \label{fig:annihilation}
\end{figure}

We now return to the $B_s$ system. By replacing the spectator
quark in the previous case, we have $\overline B_s\to D\phi$
decays replacing the role of $\overline B\to D K^{(*)}$ decays, as
depicted in Fig.~\ref{fig:DK}, in the $\gamma$ program~\cite{GL}.
Unlike the $\overline B$ case, both $\overline B_s\to D^0\phi$ and
$\overline D {}^0\phi$ modes are color suppressed decays.
Consequently, the corresponding $b\to u$ and $b\to c$ amplitude
ratio is estimated as $r_{B_s}\simeq
R_b\equiv\sqrt{\bar\rho^2+\bar\eta^2}\simeq0.4$~\cite{HFAG,PDG},
which is several times greater than $r_B$, giving a much prominent
interference effect~\cite{GL}. The $\overline B_s\to D^0\phi$
decay can be related to other decays by using the topological
approach~\cite{QD}, which is closely related to SU(3) symmetry.
Indeed the $\overline B_s\to D^0\phi$ decay is similar to other
color-suppressed modes, such as $\overline B {}^0\to
D^0\rho^0,\,D^0\omega$, as one can see by replacing $s\bar s$ and
$V_{us}$ in the second diagram of Fig.~\ref{fig:DK} by $d\bar d$
and $V_{ud}$, respectively. These modes were observed with
${\mathcal B}(\overline B {}^0\to D^0\rho^0)=(2.9\pm1.1)\times
10^{-4}$ and ${\mathcal B}(\overline B {}^0\to
D^0\omega)=(2.5\pm0.6)\times 10^{-4}$~\cite{PDG}, which are larger
than naive factorization expectations. In addition to the
color-suppressed diagram the $\overline B {}^0\to D^0\rho^0$ and
$D^0\omega$ amplitudes receive annihilation diagram contributions
(similar to the second diagram shown in
Fig.~\ref{fig:annihilation}), but with different relative signs.
The measured rates roughly satisfy ${\mathcal B}(\overline B
{}^0\to D^0\rho^0)\simeq {\mathcal B}(\overline B {}^0\to
D^0\omega)$ and, consequently, imply the sub-dominant role of the
annihilation contribution plays in these modes.
Assuming SU(3) symmetry and neglecting the annihilation
contribution, the $\overline B_s\to D^0\phi$ rate can be estimated
from these decay rates by using~\footnote{In the right-hand-side
of the equation, the annihilation amplitude only enters
quadratically. Its contribution can be safely neglected. Also note
that the $\overline B {}^0\to D^0 \rho^0(\omega)$ amplitude has an
additional factor of $1/\sqrt2$ due to the $\rho^0(\omega)$ wave
function.}
 \begin{eqnarray}
 {\mathcal B}(\overline B_s\to D^0\phi)\simeq
 \frac{ \tau_{B_s}}{\tau_{B_d}}\left| \frac{V_{us}}{ V_{ud}}\right|^2
 [{\mathcal B}(\overline B {}^0\to D^0 \rho^0)+{\mathcal B}(\overline B {}^0\to D^0 \omega)]
 \simeq 3\times 10^{-5},
 \end{eqnarray}
where $\tau_{B_d,B_s}$ are the lifetime of $B_{d,s}$ mesons with
$\tau_{B_s}/\tau_{B_u}\simeq0.95$~\cite{PDG}. Our estimation of
the $\overline B_s\to D^0\phi$ rate is one order of magnitude
larger than the previous one~\cite{ADK}. The Gronau-London method
should be useful in the extraction of $\gamma$ in the $B_s$
system.

After realizing the applicability of the Gronau-London method in
the $B_s$ system, we propose several additional theoretical clean
modes adding to the $\gamma$ program. The tree $B_s\to
D^0h^0,\,\overline D {}^0 h^0$ decays with
$h^0=\pi^0,\,\eta,\,\eta',\,\rho^0,\,\omega$, do not contain any
penguin contribution. The $B_s\to D^0\eta,\,\overline D {}^0\eta'$
modes receive contributions from color-suppress tree and
$W$-exchange diagrams as depicted in Fig.~\ref{fig:DK} and
\ref{fig:annihilation}, while others are pure weak annihilation
modes.

The $B_s\to D^0 h^0$ rates can be estimated by using the
$\overline B {}^0\to D^0 h^0$ rates in the topological amplitude
approach~\cite{QD}. We have
 \begin{eqnarray}
 \label{eq:QD}
 A(\overline B^0\to D^0 \pi^0)&=&\frac{V_{cb} V_{ud}^*}{\sqrt2} (E-C),
  \nonumber\\
 A(\overline B^0\to D^0 \eta)&=&\frac{V_{cb} V_{ud}^*}{\sqrt2} \cos\psi (E+C),
 \nonumber\\
 A(\overline B^0\to D^0 \eta')&=&\frac{V_{cb} V_{ud}^*}{\sqrt2} \sin\psi(E+C),
 \nonumber\\
 A(\overline B^0\to D_s^+ K^-)&=&V_{cb} V_{ud}^* E,
 \end{eqnarray}
 and
 \begin{eqnarray}
 A(\overline B_s\to D^0 \pi^0)&=&\frac{V_{cb} V_{us}^*}{\sqrt2} E',
 \nonumber\\
 A(\overline B_s\to D^0 \eta)&=&\frac{V_{cb} V_{us}^*}{\sqrt2} (-\sin\psi \sqrt2\,C'+\cos\psi E'),
 \nonumber\\
 A(\overline B_s\to D^0 \eta')&=&\frac{V_{cb} V_{us}^*}{\sqrt2} (\cos\psi \sqrt2\,C'+\sin\psi E'),
 \nonumber\\
 A(\overline B_s\to \overline D {}^0 \pi^0)&=&\frac{V_{ub} V_{cs}^*}{\sqrt2} E'',
 \nonumber\\
 A(\overline B_s\to \overline D {}^0 \eta)&=&\frac{V_{ub} V_{cs}^*}{\sqrt2} (-\sin\psi\,\sqrt2 C''+\cos\psi E''),
 \nonumber\\
 A(\overline B_s\to \overline D {}^0 \eta')&=&\frac{V_{ub} V_{cs}^*}{\sqrt2} (\cos\psi\,\sqrt2 C''+\sin\psi E''),
  \end{eqnarray}
where $C,C',C''$ and $E,E',E''$ are (complex) color-suppressed and
$W$-exchange amplitudes, respectively, containing possible
final-state-interaction (FSI) effects, and $\psi=39.3^\circ$ is
the mixing angle of the $\eta$ and $\eta'$ non-strange and strange
contents~\cite{eta}
 \begin{eqnarray}
 \left(
\begin{array}{c}
\eta\\
      \eta^\prime
\end{array}
\right)= \left(
\begin{array}{cc}
\cos\psi &-\sin\psi\\
\sin\psi &\cos\psi
\end{array}
\right) \left(
\begin{array}{c}

\eta_q\\
      \eta_s
\end{array}
\right)
 \end{eqnarray}
with $\eta_q=(u\bar u+d\bar d)/\sqrt2$ and $\eta_s=s\bar s$.
The color suppressed rates are measured to be ${\mathcal
B}(\overline B {}^0\to D^0\pi^0)=(2.53\pm0.20)\times10^{-4}$,
${\mathcal B}(\overline B {}^0\to D^0\eta)=(2.11\pm0.33)\times
10^{-4}$, ${\mathcal B}(\overline B {}^0\to
D^0\eta')=(1.26\pm0.23)\times 10^{-4}$ and ${\mathcal B}(\overline
B {}^0\to D_s^+ K^-)=(3.8\pm1.3)\times 10^{-5}$~\cite{D0h0,PDG}.
These decay rates are much larger than the naive factorization
expectations. There are some theoretical efforts in understanding
the largeness of these decay
modes~\cite{D0h0theory1,D0h0theory2,D0h0theory3,CH05}.
Considering, for example, the $\overline B {}^0\to D_s^+ K^-$
decay, in the rescattering approach~\cite{CH05}. Its large rate is
feed from the color-allowed $D^+\pi^-$ one, through the
rescattering process $D^+(c\bar u)\pi^-(u \bar d)\to D_s^+(c\bar
s)K^-(s\bar u)$ with the annihilation (creation) of $u\bar u$
($s\bar s$) quark pair in the initial (final) state.

The measured $\overline B {}^0\to D^0h^0$ rates are useful in
estimating $\overline B_s\to D^0 h^0$ rates. In the SU(3) limit,
we have $C=C'$ and $E=E'$.
For $\overline B_s\to D^0\eta,D^0\eta'$ modes, we have
\begin{eqnarray}
 {\mathcal B}(\overline B_s\to D^0\eta,D^0\eta')
 &\equiv&{\mathcal B}(\overline B_s\to D^0\eta)+{\mathcal B}(\overline B_s\to
 D^0\eta')
 \nonumber\\
 &\simeq&
 \frac{\tau_{B_s}}{\tau_{B_d}}\left|\frac{V_{us}}{ V_{ud}}\right|^2
 \bigg[{\mathcal B}(\overline B {}^0\to D^0\pi^0)+{\mathcal B}(\overline B {}^0\to D^0\eta)
 \nonumber\\
 &&+{\mathcal B}(\overline B {}^0\to D^0\eta')-\frac{1}{2}{\mathcal B}(\overline B {}^0\to D_s^+K^-)\bigg]
 \nonumber\\
 &\simeq&3\times 10^{-5}.
 \end{eqnarray}
To further estimate $D^0\eta$ and $D^0\eta'$ rates, we need
information on $R\equiv E'/C'$.
Using the measured color-suppressed $B^0$ decay rates and
Eq.~(\ref{eq:QD}), it is straightforward to obtain the best fitted
value of $E/C=0.26\, e^{\pm i 72^\circ}$. By assuming $R(\equiv
E'/C')\simeq E/C$ under SU(3), we estimate
 \begin{eqnarray}
 {\mathcal B}(\overline B_s\to D^0\eta)\simeq
 {\mathcal B}(\overline B_s\to D^0\eta, D^0\eta')
 \frac{|-\sqrt2\sin\psi+\cos\psi R|^2}{2+|R|^2}\simeq 1\times 10^{-5},
 \nonumber\\
 {\mathcal B}(\overline B_s\to D^0\eta')\simeq
 {\mathcal B}(\overline B_s\to D^0\eta, D^0\eta') \frac{|\sqrt2\cos\psi+\sin\psi R|^2}{2+|R|^2}
 \simeq 2\times 10^{-5},
 \end{eqnarray}
which are of the same order as ${\mathcal B}(\overline B_s\to
D^0\phi)$.

%

The pure $W$-exchange $\overline B_s\to D^0\pi^0$ decay rate can
be estimated in a similar manner as
 \begin{eqnarray}
 {\mathcal B}(\overline B_s\to D^0\pi^0)\simeq
 \frac{\tau_{B_s}}{\tau_{B_d}}\left|\frac{V_{us}}{\sqrt2 V_{ud}}\right|^2
 {\mathcal B}(\overline B {}^0\to D_s^+K^-)\simeq 1\times 10^{-6}.
 \end{eqnarray}
In fact, when take into account the SU(3) breaking effects, the
$\overline B_s\to D^0\pi^0$ decay rate could be larger than the
above estimation, since unlike the $\overline B {}^0\to D_s K$
decay no creation of the $s\bar s$ pair is needed in the final
state (see Fig.~\ref{fig:annihilation}).

Note that our estimation of the $\overline B_s\to D^0\pi^0$ rate
is similar to a recent one~\cite{CF}, while our predicted
$\overline B_s\to D^0\eta,D^0\eta$ rates are smaller than theirs
by a factor of 20. This is because, the CKM factor $V_{ud}$
instead of $V_{us}$ was used in \cite{CF} for the $\overline
B_s\to D^0\eta^{(\prime)}$ amplitudes.

The extraction of $\gamma$ in $\overline B_s\to D^0h^0$ modes can
be preformed by employing the GLW~\cite{GL,GW} method. It should
be clear that other methods, such as ADS~\cite{ADS} and $DK$
Dalitz~\cite{Dalitz,DalitzBelle0304} can also be used. However, as
$r_{B_s}$ is several times greater than $r_B$, the GLW method
should be more favorable in reducing additional uncertainties.
By the standard construction, we have~\footnote{Note that an
additional negative sign in the last equation is due to the $CP$
quantum number of $h^0$ and a $(-)^L$ factor, where $L$ is the
orbital angular momentum.}
 \begin{eqnarray}
 A(\overline B_s\to D^0h^0)&=&a,
 \nonumber\\
 A(\overline B_s\to \overline D {}^0h^0)&=&b e^{-i\gamma}e^{i\delta},
 \nonumber\\
 \sqrt{2}A(\overline B_s\to D_{CP\pm}h^0)&=&(a \pm b e^{-i\gamma} e^{i\delta}),
 \nonumber\\
  \sqrt{2}A(B_s\to D_{CP\pm}h^0)&=&\mp(a \pm b e^{i\gamma}
 e^{i\delta}),
 \end{eqnarray}
where $D_{CP\pm}$ are defined as $(D^0\pm \overline D ^0)/\sqrt2$,
$a$, $b$ are real numbers with suitable phase convention and
$\delta$ is the strong phase difference.
All four unknowns $\gamma$, $a$, $b$, $\delta$ can be obtained by
measuring the four tagged $\overline B_s\to D_{CP\pm}h^0$ and $
B_s\to D_{CP\pm}h^0$ decay rates. It is useful to define~\cite{GL}
 \begin{eqnarray}
 A_{\pm}&\equiv&
 \frac{\Gamma(\overline B_s\to D_{CP\pm}h^0)-\Gamma(B_s\to D_{CP\pm}h^0)}
      {\Gamma(\overline B_s\to D_{CP\pm}h^0)+\Gamma(B_s\to D_{CP\pm}h^0)}
 =\frac{\pm2 r_{B_s}\sin\gamma\sin\delta}
       {1+r_{B_s}^2\pm 2 r_{B_s}\cos\gamma\cos\delta},
 \nonumber\\
 R_{\pm}&\equiv&
 \frac{\Gamma(\overline B_s\to D_{CP\pm}h^0)+\Gamma(B_s\to D_{CP\pm}h^0)}
      {\Gamma(\overline B_s\to D^0 h^0)+\Gamma(B_s\to D^0h^0)}
 =\frac{1+r_{B_s}^2\pm 2 r_{B_s} \cos\gamma\cos\delta}
       {1+r_{B_s}^2},
 \end{eqnarray}
where $r_{B_s}\simeq R_b\simeq 0.4$. It should be noted that the
measurement of the asymmetry $A_\pm$ requires tagging, while the
measurement of $R_\pm$ is untagged.
In \cite{HL}, weak annihilation modes of $B_s\to D^\pm\pi^\mp$
having rate similar to ${\mathcal B}(B_s\to D^0\pi^0,\overline D
{}^0\pi^0)$ were proposed for extracting $\gamma$. However,
contrary to our case, time-dependent tagged rates are
necessary~\cite{HL}.

As a result of the same topological amplitudes for $b\to u$ and
$b\to c$ transitions, the strong phase difference $\delta$ is
likely to be small. In this case, a large $r_{B_s}$ value does not
necessary lead to a large $CP$-asymmetry $A_{\pm}$, but it is
still very useful in producing the interference effects in the
$D_{CP\pm} h^0$ rates. For illustration, using $\delta=0$,
$r_{B_s}=0.4$ and $\gamma=60^\circ$, we obtain
 \begin{eqnarray}
 R_{+}=1.34,\quad R_{-}=0.66.
 \end{eqnarray}
The measurements of $R_{\pm}$ provide $\gamma$ and $r_{B_s}$
values. The vanishing strong phase approximation is useful in
extracting or constraining $\gamma$ using less data. It can be
verified by measuring $A_\pm$, when more data is available. Since
the $b\to u$ and $b\to c$ amplitudes are of similar size, the
direct $CP$ asymmetry will be very sensitive to the strong phase
difference. In fact, similar arguments also apply to $B^0\to D^0
K^0, \overline D {}^0 K^0$ decays. The measurement of direct $CP$
violation in $B^0\to D_{CP} K^0$ decays, will provide the
information of the usefulness of the vanishing strong phase
approximation.

It is interesting to give the $\delta=0$ argument in the
rescattering picture. For example, as in the $\overline B {}^0\to
D_s^+K^-$ case, the $\overline B_s\to D^0\pi^0(\overline D
{}^0\pi^0)$ rate is mainly feed from the color-allowed $D^+_s K^-
(D_s^- K^+)$ one, through the rescattering $D^+_s(c\bar s)K^-(s
\bar u)\to D^0(c\bar u)\pi^0(u\bar u)\, [D^-_s(\bar c s)K^+(\bar s
u)\to \overline D^0(\bar c u)\pi^0(\bar u u)]$ with the
annihilation and creation of $s \bar s$ and $u\bar u$ quark pair
in the initial and final states, respectively~\cite{CH05}. The
tree-allowed $D_s^\pm K^\mp$ amplitudes do not have any strong
phase difference, while the $D^+_s(c\bar s)K^-(s \bar u)\to
D^0(c\bar u)\pi^0(u\bar u)$ and $D^-_s(\bar c s)K^+(\bar s u)\to
\overline D^0(\bar c u)\pi^0(\bar u u)$ annihilation rescattering
amplitudes are related by charge conjugation, which is respected
by strong interactions. Consequently, the strong phase difference
in $\overline B_s\to D^0\pi^0$ and $\overline D {}^0 \pi^0$
amplitudes should be small. The above consideration also applies
to other modes, including those with $C',C''$, as long as they are
long distant dominated (as hinted by the $\overline B {}^0\to
D^0h^0$ data).
For the case of $D_{CP}V$, the amplitudes $C'$ and $C''$, $E'$ and
$E''$ can be different in signs~\cite{Chiang}, but we do not
expect a large strong phase difference.

In conclusion, we point out that the large enhancement in
color-suppress decay rates observed in $\overline B$ decays
suggest similar enhancement in the color-suppress $B_s$ decay
rates. The GLW method in extracting $\gamma$ using $B_s\to
D^0\phi,\,\overline D {}^0\phi$ is not limited to the color
suppressed decay modes as previously believed.
We also suggest several new theoretical clean modes in the
extraction of $\gamma$ in $B_s$ decays. These modes are
color-suppressed $B_s\to D^0 h^0,\,\overline D {}^0 h^0$ decays,
with $h^0=\pi^0,\,\eta,\,\eta',\,\rho^0,\,\omega$, in addition to
the $h^0=\phi$ case. They are free of penguin contributions.
The extraction of $\gamma$ can be performed as in the $D_{CP}\phi$
case. These $D^0h^0$ rates are of order $10^{-6}\sim 10^{-5}$.
A combined analysis could be useful in reducing the statistical
uncertainties in the $\gamma$ extraction.
No information on the $B_s$--$\overline B_s$ mixing is required.
While the mixing is sensitive to New Physics, the $\gamma$
extraction in this case is expected to be insensitive to NP and
does not require a $\Delta m_{B_s}$ value as predicted by the
standard model. It can be considered as a complementary to the
$D^\pm_s K^\mp$ method.
The $r_{B_s}$ value is expected to be $R_b\simeq 0.4$, while the
strong phase difference between $b\to u$ and $b\to c$ amplitudes,
both are of the same topological types, are likely to be small. In
this case, the $CP$ asymmetries are suppressed and the untagged
measurements will provide very useful information in the
extraction of $\gamma$.

\begin{acknowledgments}
 The author would like to thank Hsiang-nan Li, Cai-Dian Lu and Hai-Yang Cheng for useful discussions.
This work is supported by the National Science Council of R.O.C.
under Grants NSC-94-2811-M-001-059 and NSC-93-2112-M-001-016.
\end{acknowledgments}

\end{document}